\documentclass[preprint,aps,showpacs,floatfix]{revtex4}

\usepackage{graphicx}
\usepackage{epsfig}
\usepackage{color}

\newcommand{\vvn}{\mathbf{v}_n}
\newcommand{\vvs}{\mathbf{v}_s}

\begin{document}


\definecolor{AsGreen}{rgb}{0.3,0.8,0.3}
\definecolor{AsRed}{rgb}{0.6,0.0,0.0}
\definecolor{AsLightBlue}{rgb}{0,0.6,0.6}
\definecolor{AsPurple}{rgb}{0.6,0,0.6}

\title{The saturation of decaying counterflow turbulence in helium~II}
\author{Michele Sciacca$^{1,2}$, Yuri A. Sergeev$^3$, Carlo F. Barenghi$^2$ and Ladik Skrbek$^4$}

\affiliation{
$^1$ Dipartimento di Metodi e Modelli Matematici,
Universit\`a di Palermo,
Viale delle Scienze, 90128 Palermo, Italy}
\affiliation{
$^2$ School of Mathematics and Statistics,
Newcastle University, Newcastle upon Tyne, NE1 7RU, England, UK}\email{c.f.barenghi@ncl.ac.uk}
\affiliation{
$^3$ School of Mechanical and Systems Engineering,
Newcastle University, Newcastle upon Tyne, NE1 7RU, England, UK}
\affiliation{
$^4$ Faculty of Mathematics and Physics, Charles University, Ke Karlovu 3, 121 16 Prague, Czech Republic}

\date{\today}
\begin{abstract}
We are concerned with the problem of the decay of a tangle of quantized vortices in He II generated by a heat current. Direct application of Vinen's equation yields the temporal  scaling of vortex line density $L \sim t^{-1}$. Schwarz and Rozen [Phys. Rev. Lett. {\bf 66}, 1898 (1991); Phys. Rev. B {\bf 44}, 7563 (1991)] observed a faster decay followed by a slower decay. More recently, Skrbek and collaborators [Phys. Rev. E {\bf 67}, 047302 (2003)]  found an initial transient followed by the same classical $t^{-3/2}$ scaling observed in the decay of grid-generated turbulence. We present a simple theoretical model which, we argue, contains the essential physical ingredients, and accounts for these apparently contradictory results.
\end{abstract}
\pacs{
67.25.dk, 
47.37.+q, 
47.27.Gs  
}
\keywords{quantum turbulence, vortices, counterflow, helium~II}
\maketitle
\section{Introduction and motivation}
\label{sec:introduction}

One of the most important problems of fluid dynamics is the decay of homogeneous isotropic turbulence \cite{Davidson,Frisch}. The problem is particularly interesting in He~II (the low temperature phase of liquid $^4$He) because it involves aspects of both classical and
quantum physics \cite{Donnelly}. Helium~II is a quantum fluid, which can be described phenomenologically, as the mixture of two fluid components: the inviscid
superfluid of density $\rho_s$ and velocity ${\bf v}_s$ and the viscous normal fluid of density $\rho_n$ and velocity ${\bf v}_n$.
Helium's total density is $\rho=\rho_n+\rho_s$. Because of quantum mechanical constraints, superfluid vorticity is concentrated into thin vortex filaments of quantized (fixed) circulation $\kappa=9.97 \times 10^{-4}\rm cm^2/s$. Superfluid turbulence consists of a tangle of such  vortex filaments. The intensity of the turbulence is described by the vortex line density $L$ (defined as the total vortex length per unit volume); in experiments, the vortex line density can be inferred using various methods (e.g. measuring temperature gradients, the attenuation of second sound, or the number of ions trapped in the quantized vortices). Superfluid turbulence can be easily generated by forcing a sample of helium~II with moving grids or propellers - techniques which also induce turbulence in classical ordinary fluids. In this paper we are concerned with thermal counterflow: this is a technique without direct analogy in classical fluid mechanics, in which turbulence is caused by the flo!
 w of heat in a two-fluids system.  The importance of counterflow turbulence arises from the cryogenic engineering applications of liquid helium, because turbulence limits the ability of helium~II to transport heat.

The counterflow configuration consists of a channel of cross section $A$ which is closed at one end and open to the helium bath at the other end. A heater (electrical resistor)  dissipates a known power $W$ at the closed end. Normal fluid carries the heat flux ${\dot Q}=W/A=\rho S T V_n$ away from the resistor (where $T$ is the temperature, $S$ the specific entropy and $V_n$ the normal fluid velocity (averaged over the cross section of the channel); at the same time superfluid flows in the other direction to conserve mass, so the total mass flux is zero
\begin{equation}
\rho_n V_n+\rho_s V_s=0,
\label{eq:counterflow}
\end{equation}
\noindent
where $V_s$ is the average of the superfluid profile. In this way a relative velocity $V_{ns}=V_n-V_s$ between normal fluid and superfluid
components is set up which is proportional to the applied heat flux, $V_{ns}={\dot Q}/(\rho_s S T)$. The pioneering experiments of Vinen \cite{Vinen-counterflow} determined that, for $V_{ns}>V_1$ (where $V_1$ is a critical velocity) a tangle of vortex lines of density $L=\gamma^2 V_{ns}^2$ fills the channel, where the coefficient $\gamma$ depends on $T$. Vinen also wrote a model equation for $dL/dt$ which balances growth and decay terms, whose steady state solution (when $dL/dt=0$) has indeed the observed form $L \sim V_{ns}^2$. The analysis of Schwarz \cite{Schwarz}, who pioneered numerical calculations of superfluid turbulence, confirmed the validity of Vinen's equation,
which he re-derived from the microscopic laws of vortex dynamics under a number of approximations.

Further experimental work by Tough \cite{Tough} discovered the existence of two regimes of superfluid turbulence: a weak regime (called T1) for $V_1<V_{ns}<V_2$ characterized by a small value of $\gamma$, and a strong regime (called T2) for $V_{ns}>V_2$ in which $\gamma$ is larger, where $V_2$ is a second critical velocity. The nature of the T1 and T2 states is not clear, particularly because the transition seems to depend on the channel's geometry;
a stability analysis \cite{Melotte} suggests that, in cylindrical channels and for $V_{ns}>V_2$, the normal fluid should become turbulent too.

The decay of counterflow turbulence was experimentally investigated by Schwarz and collaborators \cite{Schwarz-Rozen,Milliken-Schwarz-PRL48}, who monitored the vortex line density $L$ as a function of time after the heater has been switched off. They discovered that an initial rapid decay of $L$ is followed by a slower decay. This result seemed puzzling, as Vinen's equation implies a decay of the form $L \sim t^{-1}$. In order to make sense of their own experiments, Schwarz and Rozen \cite{Schwarz-Rozen} developed a phenomenological model which includes the normal fluid's viscous dissipation, and managed to fit the model to the experimental data.

In the years that followed the attention moved away from thermal counterflow and turned to other forms of turbulence.
Donnelly and collaborators \cite{Smith-grid,Stalp} found that the decay of the vortex line density behind a towed grid follows a
$L \sim t^{-3/2}$ behaviour, which is consistent with the classical decay of large turbulent eddies \cite{Vinen-Niemela}.  The result stimulated the study of similarities between superfluid turbulence and classical turbulence, and researchers observed the same $k^{-5/3}$ Kolmogorov energy spectra in continuously forced turbulence \cite{Tabeling}, the same pressure drops along pipes and channels \cite{Walstrom}, and even the same drag crisis for a rapidly moving sphere \cite{Smith-sphere}.

More than ten years after Schwarz and Rozen \cite{Schwarz-Rozen}, Skrbek and collaborators performed more experiments on the decay of counterflow turbulence in Prague \cite{Skrbek, QFS}. They found that, after an initial transient which seems to depend on the applied heat flux, the vortex line density decays as $L \sim t^{-3/2}$, the same decay observed in the towed grid experiment, in apparent disagreement with Vinen's equation and Schwarz {\it et al.} \cite{Milliken-Schwarz-PRL48, Schwarz-Rozen}. The Prague results excited the low temperature physics community, because they strongly suggested that the decay of counterflow turbulence (which up to that time was generally considered a different, non-classical form of turbulence) could be brought into the domain of helium~II experiments which can be understood in terms of classical fluid dynamics. But the disagreement between Prague classical $t^{-3/2}$ behaviour \cite{Skrbek, QFS} and the slower decay
observed by Schwarz and Rozen \cite{Schwarz-Rozen} stood out without any serious attempts to explain it.

The aim of this paper is to reconcile the experiments of Schwarz and collaborators \cite{Milliken-Schwarz-PRL48,Schwarz-Rozen}
with those performed in Prague \cite{Skrbek} and show that they are not in contradiction with each other. To achieve this aim we shall need to correct the original model of Schwarz and Rozen \cite{Schwarz-Rozen}. The rest of the article is organized as follows. Section \ref{sec:vinen} introduces and discusses Vinen's equation. In Section \ref{sec:experimets} we describe the relevant experiments, while the following Section \ref{sec:model} introduces our model, with results given in Section \ref{sec:results}. We discuss them and draw conclusions in Section\ref{sec:discussion}.

\section{Vinen's equation}
\label{sec:vinen}

In this section we review the basics of superfluid turbulence in counterflow experiments. We make use of the evolution equation for the vortex line density first introduced by Vinen \cite{Vinen-counterflow} and then microscopically derived by Schwarz \cite{Schwarz} using the Local Induction Approximation \cite{Saffman}.

In Schwarz's notation, the space curve ${\bf s}={\bf s}(\xi,t)$ is the position vector along a vortex line, where $\xi$ is arc length; if we denote the derivative with respect to arc length with a prime, then ${\bf s}'$ is the unit tangent vector, ${\bf s}''$ is along the normal direction, $1/\vert {\bf s}'' \vert$ is the local radius of curvature and ${\bf s}'\times {\bf s}''$ points along the binormal direction.
Since the vortices tend to grow in the plane perpendicular to the direction of the counterflow velocity ${\bf v}_{ns}={\bf v}_n-{\bf v}_s$
(Donnelly-Glaberson instability \cite{Glaberson,Cheng-Cromar-Donnelly}), the direction which is binormal to the vortices, ${\bf s}'\times{\bf s}''$, tends to be parallel to the direction of the counterflow, ${\bf v}_{ns}$.

According to Schwarz's calculation, the vortex line density equation near equilibrium obeys
\begin{equation}
\frac{dL}{dt}=\alpha \left( {\bf v}_{ns} \cdot {\bf I}_l L^{3/2}-\beta c_2^2 L^2 \right),
\label{eq:vinen}
\end{equation}
where $\alpha$ is a temperature dependent mutual friction coefficient related to Vinen's mutual friction coefficient $B$ \cite{BDV,Donnelly-Barenghi} by $\alpha=B\rho_n/(2 \rho)$,
\begin{equation}
I_{\ell} {\hat {\bf v}_{ns}}=L^{-1/2}<{\bf s}'\times {\bf s}''> \quad {\rm and} \quad <|{\bf s}''|^2>=c_2^2 L,
\end{equation}
${\hat {\bf v}_{ns}}\|{\bf v}_{ns}$ is the unit vector   and $<f>=\int f(\xi) d\xi/(\Omega L)$ denotes the line-length weighted average of any quantity $f$ where $\Omega$ is the volume and the integral is along the vortex lines. The coefficient $\beta$ is the self-induction parameter, defined by
\begin{equation}
\beta=\frac{\kappa}{4 \pi}\ln{(L^{-1/2}/a_0)},
\label{eq:beta}
\end{equation}
\noindent
where $a_0 \approx 10^{-8}\rm~cm$ is the vortex core radius.

A number of authors have dealt with Vinen's equation (\ref{eq:vinen}) and have attempted to include more physical ingredients, taking into account inhomogeneity, anisotropy, rotation  and/or boundaries. These extensions were made essentially using dimensional analysis, because of the difficulty in treating these problems from a microscopical viewpoint \cite{Mongiovi-JP17-2005,Mongiovi-PRB75-2007}. The main physical ingredients which should affect counterflow turbulence are anisotropy and the presence of the walls. The former appears in the first term at the RHS of Eq.~(\ref{eq:vinen}), which, according to the Vinen's interpretation, induces growth of the vortex line length. Since this paper is concerned with vortex line decay, this term is important only in setting up the steady state from which the decay begins, hence it seems sufficient for our purposes without any further modifications. In a more exhaustive analysis which is not immediately relevant to our aim, the first !
 term should be substituted by a tensor, as in Ref~\cite{Mongiovi-submitted-2010}, because the growth of the vortex line density occurs not only in the direction of the counterflow velocity.

The channel walls also affect the evolution of the vortex line density \cite{Martin-PRB27-1983}. Mongiov\'{i} and collaborators
\cite{Mongiovi-JP17-2005,Mongiovi-PRB71-2005} extended Vinen's equation (\ref{eq:vinen}) to include boundaries. The relevant dimensionless parameter is the ratio $\ell/D$, where $\ell\simeq L^{-1/2}$ is the average inter vortex spacing and $D$ is the channel size. They found ~\cite{Mongiovi-JP17-2005} that for small times the influence of the walls is not important and $L$ decays as $t^{-1}$, whereas  for long time (in the dilute vortex tangle limit) a term which is exponential in time appears. The experimental data which we consider \cite{Schwarz-Rozen,Skrbek} are not in this regime: Skrbek \cite{Skrbek} at $T=1.6~\rm K$, $\dot{Q}/A=0.08~\rm W/cm^2$ and channel's diameter $D=0.9~\rm cm$ has $\ell=7.4\times 10^{-3}~\rm cm$ at $t=0$ (see Table \ref{tableSkr}), and $\ell=0.031~\rm cm$ at $t=10~\rm s$;
Schwarz and Rozen \cite{Schwarz-Rozen} at $T=1.9~\rm K$, $\dot{Q}/A=0.108~\rm W/cm^2$ and channel's small size $D=1~\rm cm$ have
$\ell=7.69\times 10^{-3}~\rm cm$ at $t=0$ (see Table \ref{tableSch}), and $\ell=0.3$ at $t=600~\rm s$. We conclude that $\ell<D$ in the range of interest, hence, for the sake of simplicity, we need not to include wall effects in Vinen's equation

The steady vortex line density solution of Vinen's equation (\ref{eq:vinen}) is
\begin{equation}
L=\gamma^2 V_{ns}^2,
\label{eq:steady}
\end{equation}
\noindent
where $\gamma=c_L/\beta$ (with $c_L=I_l/c_2^2$) is a coefficient which depends on the temperature, and, in some cases, on whether the turbulence is in the so-called T1 or T2 state \cite{Tough}. We shall discuss the issue of $\gamma$ in Section \ref{sec:discussion}.

\begin{table}
  \centering
\begin{tabular}{|c|c|c|c|c|c|c|c|}
  \hline
  $T$ & $\dot{Q}/A$ & $V_{ns}(0)$ & $\gamma$ & $L(0)$ $\times 10^{5}$  & $c_{L}$ &  $I_{l}$ & $L(20)$ \\ \hline
  1.6 & 0.88 & 16.012 & 93 & 22.176 & 0.082 &  0.754 & 47  \\ \hline
  1.6 & 0.57 & 10.372 & 93 & 9.3 & 0.085 &  0.727 & 50 \\  \hline
  1.6 & 0.31 & 5.641 & 93 & 2.75 & 0.09 &  0.687 &  56 \\  \hline
  1.6 & 0.22 & 4.003 & 93 & 1.386 & 0.092 &  0.672 &  59 \\ \hline
  1.6 & 0.14 & 2.547 & 93 & 0.561  & 0.095 &  0.651 & 63 \\ \hline
  1.6 & 0.08 & 1.456 & 93 & 0.183 & 0.099 &   0.625 &  70 \\
  \hline
\end{tabular}
  \caption{Parameters from the experiments performed by Skrbek {\it et al.} \cite{Skrbek} for different heat fluxes:
Temperature $T$ (K), heat flux $\dot{Q}/A$ ($\rm W/cm^2$), steady state counterflow velocity $V_{ns}(0)$ ($\rm cm^2/s$),
$\gamma$ ($\rm s/cm^2$), initial vortex line length $L(0)$ ($\rm 1/cm^2$), and parameters $c_L$ and $I_l$. In the last column we write the value of $L$ at $t=20\, \rm sec$ from our simulations. }
\label{tableSkr}
\end{table}

\begin{table}
  \centering
\begin{tabular}{|c|c|c|c|c|c|c|c|}
  \hline
  $T$ & $\dot{Q}/A$ & $V_{ns}(0)$ & $\gamma$ & $L(0)$  $\times 10^{5}$ & $c_{L}$ & $I_{l}$ & $L(1000)$  \\ \hline
  1.9 & 0.065 & 0.558 & 140.1 & 0.061 &   0.156 & 0.507 & 1.17 \\ \hline
  1.9 & 0.108 & 0.928 & 140.1 & 0.169 &  0.15 &  0.525 & 0.76 \\ \hline
  1.9 & 0.237 & 1.778 & 140.1 & 0.62 &  0.143 & 0.551 & 1.03 \\ \hline
  \hline
\end{tabular}
  \caption{Parameters from the experiments performed by Schwarz and Rozen \cite{Schwarz-Rozen} at $T=1.9\, \rm K$. The table lists temperature $T$ (K), heat flux $\dot{Q}/A$ ($\rm W/cm^2$), steady state counterflow velocity $V_{ns}(0)$ ($\rm cm^2/s$),
$\gamma$ ($\rm s/cm^2$), initial vortex line length $L(0)$ ($\rm 1/cm^2$), and parameters $c_L$ and $I_l$. The values of $\gamma$ are taken from Ref.\,\cite{Tsubota-arXiv-2010}, and the values of $c_L$ and $I_l$ are found from Ref.\,\cite{Schwarz-Rozen}. In the last column we write the value of $L$ at $t=1000\, \rm sec$ from our simulations. }
\label{tableSch}
\end{table}

It is generally assumed that in counterflow turbulence the vortex line density is relatively homogeneous, has only a mild anisotropy, is rather featureless (unlike ordinary turbulence, where the energy is distributed over the length scales according to the Kolmogorov $-5/3$ law), and that the only characteristic length scale of the tangle is $\ell$; this means that both the average radius of curvature and the average distance between the vortex lines are of the order of magnitude of $\ell$. According to Vinen's interpretation, the first term
of equation (\ref{eq:vinen}) describes to the generation of vortex line due to the relative velocity between normal and superfluid component, and the second term models the destruction of vortex lines. Essentially, the counterflow supplies energy for the growth of vortex length through the first term of Vinen's equation and the tangle converts back its length into energy. In the steady state situation, two contributions balance each other, and formula (\ref{eq:steady}) holds true.

In order to study the decay of the vortex line density in counterflow experiments we  need dynamical equations for the superfluid helium.
The most commonly used model is the two fluids model \cite{Donnelly}.
The two-fluid equations are

\begin{eqnarray}
\rho_s \left(
\frac{\partial \vvs}{\partial t} + \vvs \cdot \nabla \vvs \right)
=-\frac{\rho_s}{\rho} \nabla p + \rho_s S \nabla T  +{\bf F}_{ns},\nonumber\\
\rho_n\left(
\frac{\partial \vvn}{\partial t} + \vvn \cdot \nabla \vvn \right)=\nonumber \\
=-\frac{\rho_n}{\rho} \nabla p - \rho_s S \nabla T
+ \mu  \nabla^2 \vvn - {\bf F}_{ns}, \nonumber\\
\nabla \cdot \vvn=0, \qquad \qquad \nabla \cdot \vvs=0,
\label{eq:cond}
\end{eqnarray}
where $p$ is the pressure and  $\mu$ the viscosity.
We do not know the precise form of the mutual friction force ${\bf F}_{ns}$,
but, for $V_{ns}>V_1$, it is reasonable to assume the
Gorter-Mellink form:
${\bf F}_{ns}\approx \rho_s \kappa \alpha L (\vvn - \vvs)\,$.

\section{Experimental data}
\label{sec:experimets}

Before starting our analysis, we summarize the necessary helium parameters  (for $\rho_s$, $\rho_n$, $S$ and $\mu$ at the temperature $T$ of a considered experiment see Table \ref{tableHe}) and the available experimental data (Tables \ref{tableSkr} and \ref{tableSch}).

\begin{table}
  \centering
\begin{tabular}{| c || c | c | c | c | c  | c |}
\hline
$T$     &  $\rho_n$      &  $\rho_s$      &  $\rho$        &  $S$  &  $\mu$   &   $\alpha$   \\ \hline
$\rm K$ &  $\rm g/cm^3$  &  $\rm g/cm^3$  &  $\rm g/cm^3$  &   $\rm cm^2/Ks^2$   &    $\rm g/s cm$   &             \\ \hline \hline
$1.6$   &  0.02358  &   0.12163  &  0.14521  &  $2824 \times 10^{3}$  & $1.306\times 10^{-5}$  &  0.097   \\ \hline
$1.9$   &  0.06103   &  0.08444  & 0.14547  &  $7255 \times 10^{3}$  & $1.347\times 10^{-5}$  &  0.206   \\
\hline
\end{tabular}
\caption{Values of selected parameters in superfluid helium used in our model.}
\label{tableHe}
\end{table}

The Prague experiments \cite{Skrbek} were performed in a circular channel ($9~\rm cm$ in diameter and $13~\rm cm$ long) at the same
temperature $T=1.6~\rm K$ and at different powers $\dot{Q}/A=0.88$, $0.57$, $0.31$, $0.22$, $0.14$ and $0.08 ~\rm W/cm^2$. The main finding was that, after a transient $t_{sat}$, the vorticity $\kappa L$ decays as $t^{-3/2}$.

The experiments of Schwarz and Rozen \cite{Schwarz-Rozen} were performed in a rectangular $1.0 \times 2.32~\rm cm$ channel which was
$24~\rm cm$ long at $T=1.4~\rm K$ and $Q=0.043 \rm W/cm^2$, $T=1.6~\rm K$ and $\dot{Q}/A=0.065~\rm W/cm^2$, $T=1.9~\rm K$ and
$\dot{ Q}/A=0.065$, $0.108$ and $0.237 \rm W/cm^2$. Schwarz and Rozen presented graphs of the time evolution of the
quantity $(\beta L_m)^{-1}$ vs $t$, where $L_m$ is $L$ times a geometrical coefficient, and $\beta$ is probably taken constant,
in the range $0.0$ to $100~\rm s$, except for $T=1.9~\rm K$ and $\dot{ Q}/A=0.108~\rm mW/cm^2$, for which data are plotted up to $1000~\rm s$.
In this last experiment, Schwarz and Rozen found that the vortex line density initially decays as $L \sim t^{-1}$, but,
after a transient, the decay becomes faster.

The experiments of Milliken and Schwarz \cite{Milliken-Schwarz-PRL48} were performed in an open geometry at $T=1.33~\rm K$, $T=1.45~\rm K$ and $T=1.60~\rm K$. They did not report the heat flux, so we cannot model their results. They claimed that the decay of the vortex lines
has the form $L(t) = 4 k L(0)/(4k+\beta L(0)t)$, where $L(0)$ is the initial value of $L$, $\beta$ is constant and $k=7.5$ is a coefficient that best fits the experimental data. No data were reported for $t>50~\rm s$.

Not all information which we need in order to set up our model is available in the literature: for example, data for $L$ vs $t$ can be
read from published figures, but in some cases the precise initial value $L(0)$ is not reported, or is not clearly visible in graphs
which plot $1/L$ vs $t$. Moreover, in the Prague experiment \cite{Skrbek} the flow channel was cylindrical in shape and the planar gold-plated micropore membranes based second sound transducers made the top and bottom of another short cylinder oriented perpendicularly to the the channel axis in the middle  of its length; the channel cross-section was therefore not of exactly uniform shape and the measured steady--state vortex line density inferred from the attenuated amplitude of standing wave second sound resonance was most likely slightly affected by this geometry.

To fill such gaps we rely on equation (\ref{eq:steady}) for the steady state: given the temperature and the initial heat flux, equation (\ref{eq:steady}) enables us to recover the initial value of the vortex line density. A problem, however, arises: the parameter $\gamma$ in equation (\ref{eq:steady}) is not well-known (different values are reported in the literature, see \cite{Tough,Martin-PRB27-1983,Tsubota-arXiv-2010} and references therein), and perhaps is not even unique (in some geometries it depends on the turbulence state being T1 or T2). Since our main result does not depend strongly on this choice, we use the values of $\gamma$ recently reported in Ref.~\cite{Tsubota-arXiv-2010}, which we interpret as referring to the T1 state; experimental values are discussed in Ref.~\cite{Tough,Childers-PRL71-1993}. The choice of $\gamma$ is discussed further in Section\,\ref{sec:discussion}.

Values of $\gamma$ given in Tables \ref{tableSkr} and \ref{tableSch} are therefore taken out from Ref.\,\cite{Tsubota-arXiv-2010}, and $L(0)$ is derived by formula (\ref{eq:steady}). The exact value of $\beta$ can be found from equation~(\ref{eq:beta}), then $c_L=\beta \gamma$. The parameters $I_l$ and $I_l/(I_{\|}-c_L I_l)$ are found using the experimental data for $c_L^{2/3}(I_{\|}-c_L I_l)^{1/3}$ respectively from figure 12 and from figure 15 of Schwarz and Rozen's paper \cite{Schwarz-Rozen}. Parameters $I_l$, $I_{\|}$ and  $c_L$ are defined in Ref.~\cite{Schwarz}.  The initial counterflow velocity $V_{ns}(0)$ is found from the known relation $\dot{Q}/A=\rho_s T S V_{ns}$.

\section{Model}
\label{sec:model}

In order to bridge the Prague experiments with those of Schwarz with coworkers, we propose the following minimal model, which, we argue, captures the essential physical ingredients. Let $x$ be the direction along the channel and $y$ the direction across it. We start from the two-fluid equations (\ref{eq:cond}) and the mass conservation equation (\ref{eq:counterflow}). It is fairly certain that the normal fluid velocity profile obeys no-slip boundary conditions $\vvn=0$ at the walls of the channel, whereas the superfluid velocity can slip along it.

Our first step is to notice that in a steady state situation, of the order of magnitude, the ratio of friction forces and viscous forces
acting on the normal fluid is
\begin{equation}
r=\frac{\rm friction~force}{\rm viscous~force}
\approx \frac{\kappa \alpha \rho D^2 L}{\mu},
\label{eq:r}
\end{equation}
\noindent
where $D$ is the size of the channel. Since $r$ depends only on $L$, and not on $V_{ns}$,  we can estimate $r$ for different temperatures and  different vortex line density in the temporal interval of interest. At $t=0$, the lowest values of $L$ in Tables \ref{tableSkr} and \ref{tableSch} are
$L(0)=0.185 \times 10^{5}~\rm cm^{-2}$ for Skrbek \emph{et al.} and $L(0)=0.061  \times 10^{5}~\rm cm^{-2}$ for Schwarz \emph{et al.}, which implies that $r=1.6 \times 10^{4}$ and $r=3.14 \times 10^{4}$, respectively. At $t=10~\rm s$, the Prague experimental data
show $L \approx 1000~\rm cm^{-2}$, that is $r=873$. At $t= 600~\rm s$  in the paper of Schwarz and Rozen, $L\approx (61 \beta)^{-1}$, hence $r\simeq 57$. The last columns of Tables \ref{tableSkr} and \ref{tableSch} show the values of $L$ at larger times resulting from our calculations: we find that the values of $r$ are smaller: $r\simeq 5$ for Schwarz and $r\simeq 50$ for Skrbek.

We conclude that, because of the large value of $r$ in the time range of interest (up to the time which we shall call $t_{sat}$),
it is reasonable to assume that the normal fluid profile is uniform in $y$, which is consistent with an experiment of Awschalom \emph{et al.} \cite{Schwarz-profile}. We can therefore neglect the relatively thin boundary layer which meets the no-slip boundary conditions at $y=0$ and $y=D$. The same argument applies to the other transverse direction $z$. Thus ${\bf v}_n$ must be independent of $y$ and $z$. Assuming that the channel is long and thus the flow is independent on $x$, we conclude that ${\bf v}_n$ can be replaced by the scalar quantity $V_n$ (in the $x$ direction), and that the nonlinear term ${\bf v} \cdot \nabla {\bf v}_n$ at the LHS is zero. This gives us an equation for $dV_n/dt$. By applying the same argument to the superfluid equation in~(\ref{eq:cond}), we obtain an equation for $dV_s/dt$. The uniform profiles $V_n$ and $V_s$ can be identified with the cross-channel averaged profiles which satisfy the counterflow condition, thus they are not independent: usin!
 g Eq.~(\ref{eq:counterflow}), we reduce the two equations to a single  equation for $V_n$, or, preferably, for $V_{ns}=(\rho/\rho_s) V_n$.

In the experiments, when the heater is switched off, there are short pressure and thermal transients during which the driving pressure
and temperature gradients change rapidly and become negligible; the former is of the order of $\Delta t_P \sim Y/c_1 \sim 10^{-4}~\rm s$
where $Y$ is the length of the channel and $c_1$ the speed of first sound; the latter is of the order of $\Delta t_T \sim Y^2/\chi$ where
$\chi$ is the thermal diffusivity, and probably even smaller. We neglect both these transients. We also neglect the longer transient during which the vortex tangle depolarizes, and the vortices lose the anisotropy imposed by the counterflow: this effect has been already studied \cite{Barenghi-Skrbek} and involves a correction to $L$ of order unity. Since the anisotropy of the tangle is imposed by the counterflow, the corresponding relaxation time is of the same order of that of $V_{ns}$.

We conclude that our minimal model reduces to the two equations
\begin{eqnarray}
\frac{d V_{ns}}{dt} =- \kappa \frac{\rho}{\rho_n} \alpha L V_{ns},
\label{eq:1}\\
\frac{dL}{dt}=\alpha I_{\ell}
\left( V_{ns} L^{3/2}-\frac{\beta}{c_L}L^2\right),
\label{eq:2}
\end{eqnarray}
\noindent
where $\beta$ is given by Eq.~(\ref{eq:beta}), with initial conditions $V_{ns}(0)={\dot Q}/(\rho_s S T)$ and $L(0)=(\gamma/(\rho_s S T))^2{\dot Q}^2$.

We stress that our model differs from the model proposed by Schwarz and Rozen \cite{Schwarz-Rozen}, which consists of three equations for $dV_n/dt$, $dV_s/dt$ and $dL/dt$. Firstly, their model does not conserve mass (during the decay of the tangle the counterflow condition still applies, because the channel is closed). Secondly, in their model $\beta$ is constant. Thirdly, and more importantly, their  equation for $dV_n/dt$ contains the term $-(\mu/\rho_n) V_n/(D/15)^2$ to model the effects of viscous dissipation.  As we have seen, the viscous forces are negligible compared to the mutual friction forces in the temporal range considered in the experiments. Moreover, Schwarz and Rozen openly
state that the factor $D/15$ (rather than, say,  $D$) was chosen for the only reason that $D/15$ gives the best fit to the experimental data. By numerically solving their equations, we found that slight changes of this arbitrary factor $D/15$ produce decay curves which are inconsistent with the data.

\begin{figure}[h]
\includegraphics[angle=0,width=0.99\linewidth]{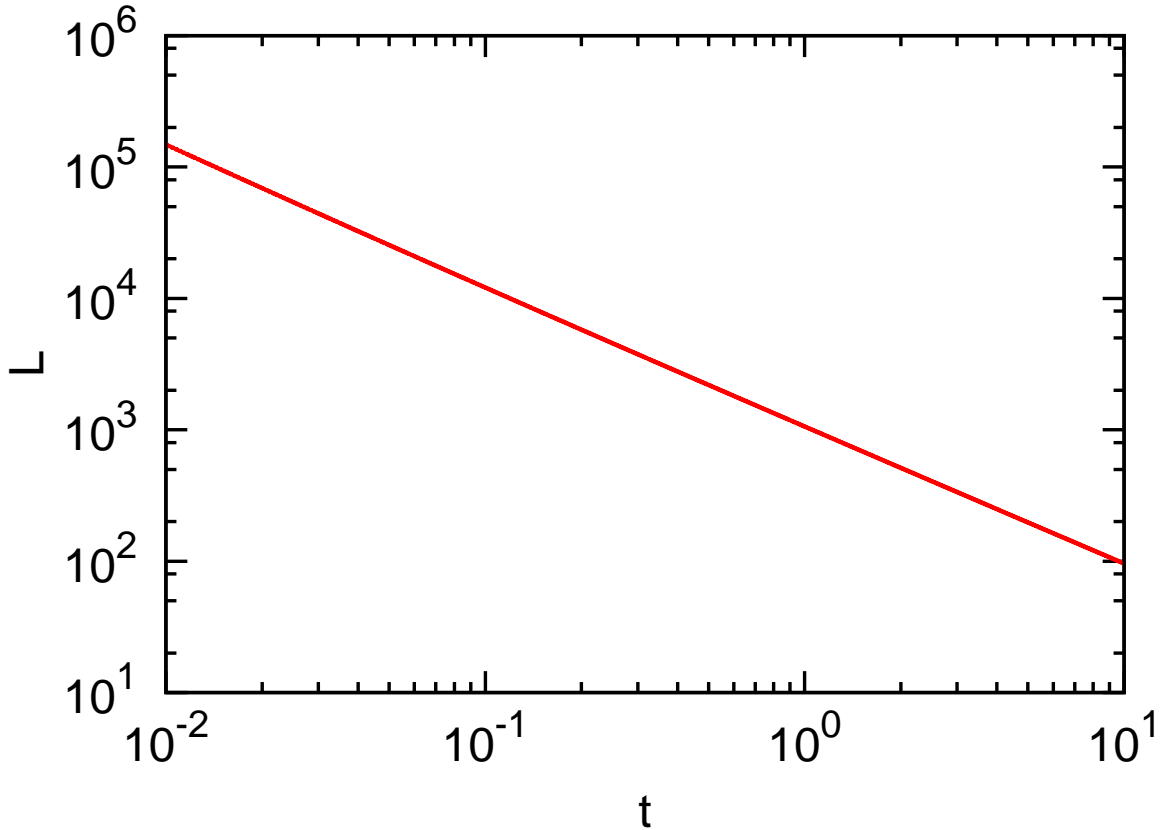}
\caption{\label{fig:1}(Color online).
Decay of vortex line density $L$ ($\rm cm^{-2}$)
vs time $t$ ($\rm s$) for the initial heat flux $\dot{Q}/A=0.88~\rm W/cm^2$
at $T=1.6~\rm K$, modelling Skrbek's experiment \cite{Skrbek}. The initial
value of the vortex line density is $22.176 \times 10^{5}\ \rm cm^{-2}$.
}
\end{figure}

\begin{figure}[h]
\includegraphics[angle=0,width=0.99\linewidth]{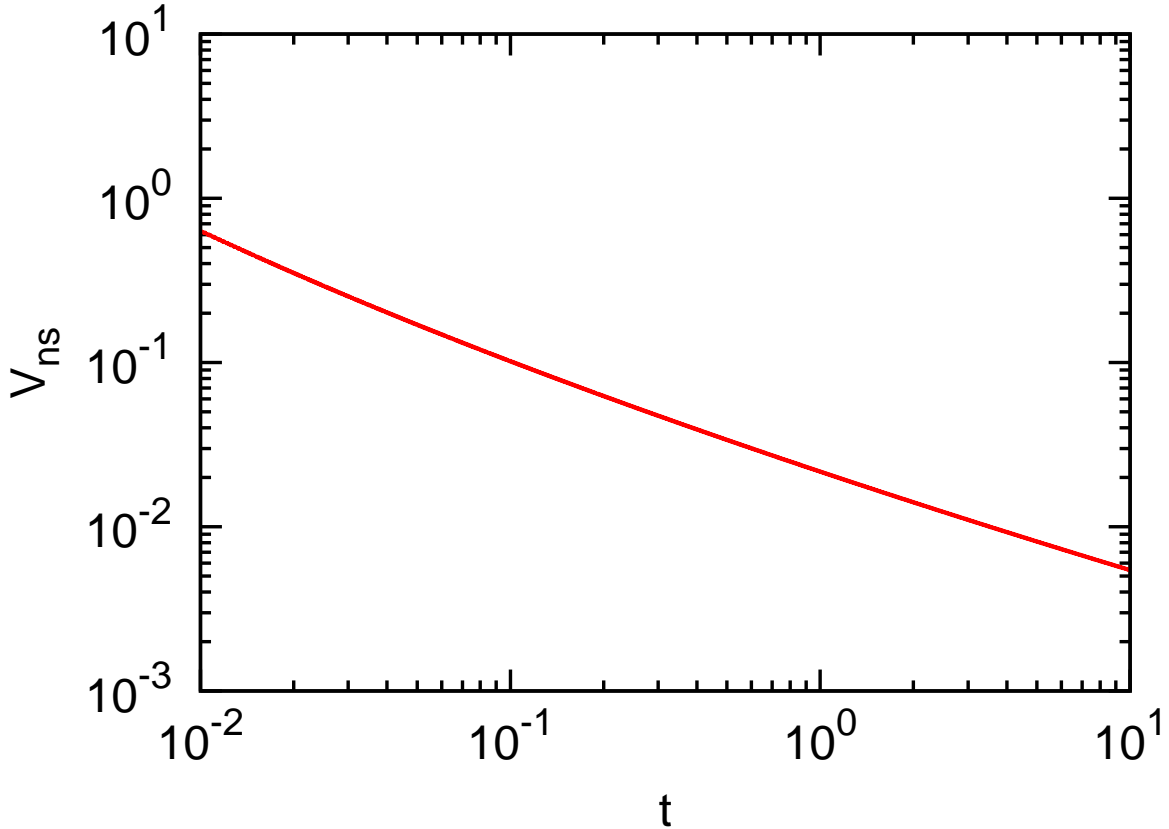}
\caption{\label{fig:2} (Color online).
Decay of the counterflow velocity $V_{ns}$ ($\rm cm/s$)
vs time $t$ ($\rm s$) for the initial counterflow turbulence
intensity corresponding to the same initial conditions of Fig.~\ref{fig:1}}.
\label{fig:2}
\end{figure}

\section{Results}
\label{sec:results}

The induction parameter $\beta$ depends on $L$ only via a logarithmic term, so in the first approximation it can be considered constant; this is what Schwarz did in his numerical simulations \cite{Schwarz} and in the model of Ref.~\cite{Schwarz-Rozen}. If $\beta$ is constant, it is apparent that Eqs.~(\ref{eq:1}) and (\ref{eq:2}) have solutions which scale as $L \sim t^{-1}$ and $V_{ns} \sim t^{-1/2}$. This means that the two terms in the bracket of Eq.~(\ref{eq:2}) remain parallel to each other (in log-log axes) and never cross: both terms scale as $t^{-2}$.

However, if $L$ changes by a large amount, which is the case in the experiments under consideration, the approximation of constant $\beta$ is not satisfactory. If $\beta$ depends on $L$ as in Eq.~(\ref{eq:beta}), Eq.~(\ref{eq:1}) and (\ref{eq:2}) cannot be solved analytically, so we integrate them numerically using the fourth order Runge Kutta method. We find that both $L(t)$ and $V_{ns}(t)$  decrease, as shown in Fig.~(\ref{fig:1}) and (\ref{fig:2}).

\begin{figure}[h]
\includegraphics[angle=0,width=0.99\linewidth]{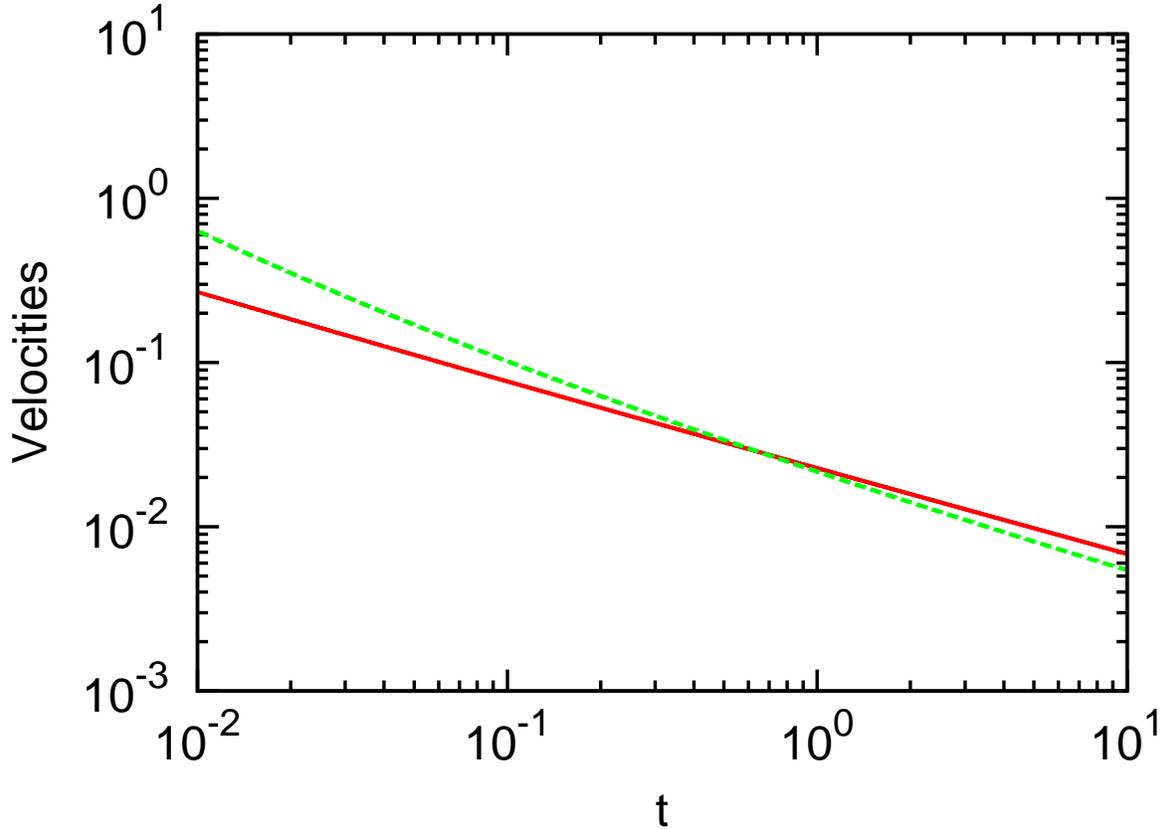}
\caption{\label{fig:3} (Color online).
Counterflow velocity $V_{ns}$ ($\rm cm/s$) (solid green line ) and
turbulent velocity $V_{\ell}$ ($\rm cm/s$) (dashed red line)
vs time $t$ ($\rm s$), corresponding to the same
initial conditions of Fig.\,\ref{fig:1}. For $c=0.7$ two curves overlap
at $t=0.655~\rm s$, which is the value of $t_{sat}$ plotted in
Fig.~\ref{fig:4} for $V_{ns}(0)=16.0124~\rm cm/s^2$.
}
\end{figure}

It is important to notice that, at a certain time $t=t_{sat}$, the counterflow velocity $V_{ns}$ becomes of the order of the typical turbulent superfluid velocity $V_{\ell}$ in the vortex tangle, which we estimate to be of the order of magnitude of $\kappa/\ell = \kappa L^{1/2}$; hereafter we set $V_{\ell}=c \kappa L^{1/2}$, where $c$ is a constant of order unity. Fig.~\ref{fig:3} shows the two curves $V_{ns}(t)$ and $V_{\ell}(t)$ near $t=t_{sat}$. Clearly for $t>t_{sat}$ the turbulence becomes qualitatively different, because the external drive $V_{ns}$ has become as weak as the background turbulence noise.

\begin{figure}[h]
\includegraphics[angle=0,width=0.99\linewidth]{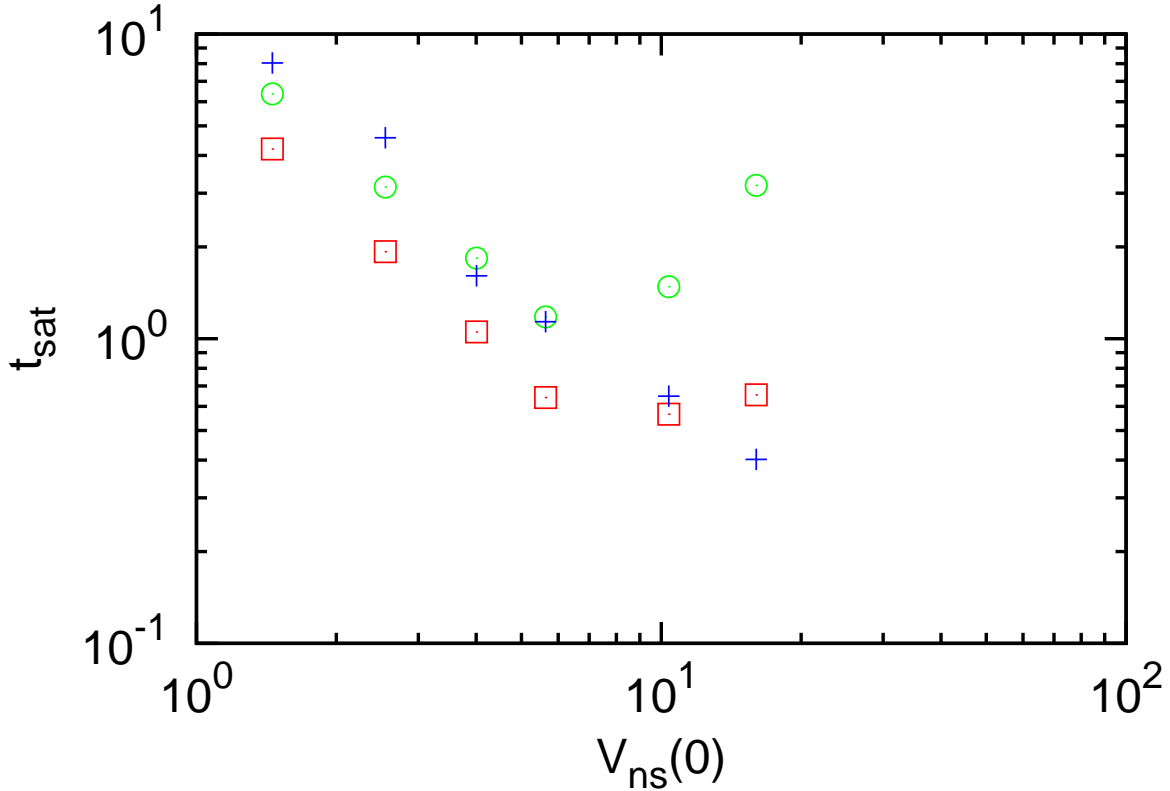}
\caption{\label{fig:4} (Color online).
Comparison of values of $t_{sat}$ ($\rm s$) vs $V_{ns}(0)$ ($\rm cm/s$) observed by
Skrbek \emph{et al.} \cite{Skrbek} (blue crosses) with those
predicted by our model with values of $t_{sat}$ for $c=0.7$ (red squares
) and $c=0.6$ (green circles).
}
\end{figure}

\begin{figure}[h]
\includegraphics[angle=0,width=0.99\linewidth]{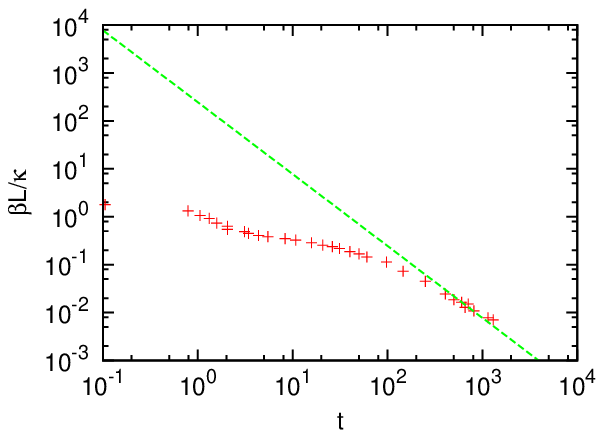}
\caption{\label{fig:5} (Color online).
Plot of $\beta L/\kappa$ vs $t$ ($\rm s$). The crosses are
experimental data from figures 17, 18, 20 measured by
Schwarz and Rozen \cite{Schwarz-Rozen} at
$T=1.9~\rm K$ and $\dot{Q}/A=0.108r~ \rm W/cmi^2$.
The (green) line shows the $t^{-3/2}$ dependence. Note the qualitative
agreement for $t>t_{sat}=600~\rm s$, as predicted by our model.
}
\end{figure}

Fig.~\ref{fig:4} shows our computed values of $t_{sat}$  for $c=0.6$ and $c=0.7$ as a function of $V_{ns}(0)$. It is apparent that there is a fairly good agreement with the Prague experiments \cite{Skrbek}.

What happens for $t>t_{sat}$ ? When the drive $V_{ns}$ has become very weak, Vinen's equation reduces to $dL/dt = -(\beta/c_L)L^2$ and one would naively infer that $L \sim t^{-1}$. A more complex scenario is thought to take place, as envisaged by various authors \cite{Smith-grid,Vinen-Niemela}, which is worth to summarise briefly.  At the beginning of the decay the superfluid velocity field is essentially random, and $\ell\approx L^{1/2}$ is the only length scale of the tangle. This means that most of the kinetic energy is concentrated at large wavenumbers, of the order of $1/\ell$. As the action of the counterflow weakens, it is reasonable to assume that this energy is shifted to smaller wavenumbers as in the decay of ordinary turbulence \cite{Corrsin}, until the energy-containing eddies become of the order of the channel's size $D$. Based on this scenario, the following model for the decay of grid turbulence in helium~II was proposed by Stalp, Skrbek and Donnelly \cite{Sta!
 lp} (see also Skrbek, Niemela and Donnelly \cite{SND} and Skrbek and Stalp \cite{SS}). If $U$ is their speed and $E$ their energy, the lifetime of these eddies is $\tau \sim D/U$ and $dE/dt \sim E/\tau \sim U^3 \sim E^{3/2}$, hence their energy decays as $E \sim t^{-2}$. If, in analogy with classical turbulence (for which the rate of energy dissipation is $\epsilon=\nu \omega^2$ where $\omega$ is the average vorticity and $\nu$ the kinematic viscosity), we identify the average superfluid vorticity with $\kappa L$, then $\epsilon=-dE/dt \sim t^{-3}$, so $\nu' \kappa^2 L^2 \sim t^{-3}$ and we conclude that $L \sim t^{-3/2}$. This is the universal decay law, which was first observed by Smith and coworkers \cite{Smith-grid} in both towed grid and counterflow experiments and later investigated in detail in Prague
experiments \cite{Skrbek, QFS} for $t>t_{sat}$.

If we plot the data of Schwarz and Rozen \cite{Schwarz-Rozen} we see in Fig.~\ref{fig:5} that, for $T=1.9~\rm K$ and $\dot{Q}/A=0.108~\rm W/cm^2$, for $t>t_{sat}=600~\rm s$ the decay is consistent with the same $t^{-3/2}$ power law as observed in Prague \cite{Skrbek, QFS}, and our calculation predicts $t_{sat}=600~\rm s$, provided that we take $c=0.77$. This agreement between the data reported in Ref.~\cite{Schwarz-Rozen} and \cite{Skrbek} in terms of time dependence has not been noticed before. It is also important to notice that the values of $t_{sat}$ are much shorter for the Prague data than for those of Schwarz and Rozen, due to the different values of vortex line density in steady state from which the decay originated.

We have considered the other two cases of Table~\ref{tableSch} at $T=1.9~\rm K$ even though Schwarz and Rozen do not plot data for long enough time to reach  $t_{sat}$ (up to $100~\rm s$). Our model predicts $t_{sat}=151.7~\rm s$ for  $\dot{Q}/A=0.065~\rm W/cm^2$. The important difference (apart from the sizes of the channels used, which are, however, nearly the same) between the Prague experiments and those of Schwarz and Rozen is thus the vortex line density, which in the latter is considerably less, hence the intersection point which defines $t_{sat}$ is not reached.

Regarding the experiment at $\dot{Q}/A=0.237~\rm W/cm^2$, we have not found an intersection $t_{sat}$ before $t=1000~\rm s$. Probably in this case our model fails because it is too simple: the small value $L=1.03~\rm cm^{-2}$ in Table~\ref{tableSch} at $t=1000~\rm s$ implies $\ell \approx 0.9~\rm cm$, which is of the same order of the small size of the channel, indeed $r\approx 5$, and viscous forces are not negligible in this case.

\section{Discussion and conclusions}
\label{sec:discussion}

The model which we propose grasps the main behaviour of Skrbek's and Schwarz and Rozen's experiments, but the exact value of the
ratio $c=V_{ns}/\kappa L^{1/2}$ cannot be determined for a number of reasons: firstly, our model is too simple to detect fluctuations of the main fields; secondly, the completeness of the Hall-Vinen-Bekarevich-Khalatnikov equation is still an open question
\cite{Mongiovi-submitted-2010}; thirdly, some parameters and initial data are not fully known; fourthly, we have neglected any transient in the first part of the decay arising from any diffusion process, any decay of normal fluid turbulence, and any possible decay from a T2 state into a T1 state.

Let us consider the state of the superfluid turbulence more carefully.  Tough and collaborators \cite{Tough} performed many experiments on counterflow in tubes, using different temperatures, heat fluxes, tube sizes, tube shapes (circular and rectangular sections) and different materials, as described in Ref.~\cite{Martin-PRB27-1983} and references therein. These studies showed that the values of $\gamma$ in (\ref{eq:steady}) depends not only on $T$ but also on the sizes of the channel and on the heat flux $\dot Q$ applied to the sample.
In circular tube there are two states (called $T1$ or $T2$) characterised by the value of $\gamma$. Since (roughly) $\gamma_{T2}\approx 2\gamma_{T1}$, the T2 state corresponds to a larger vortex line density (more intense turbulence). The nature of the difference between the T1 state and the T2 state is not clear, although it has been suggested \cite{Melotte} that the transition at $V_{ns}=V_2$ from T1 to T2
corresponds to a transition to turbulence in the normal fluid (which would enhance the intensity of the superfluid vortex tangle).

Prague experiments were performed in a circular pipe, whose diameter ($0.9~\rm cm$) was much larger than diameters used by Tough and collaborators. If we use Martin and Tough's paper \cite{Martin-PRB27-1983} to determinate the critical velocity  $V_2$ as a function of $D$ and $T$, we find $V_{2}=0.244~\rm cm/s$, which is smaller than any initial velocity in Prague experiments (see Table~\ref{tableSkr}). We are thus led to suspect that in Prague experiments the turbulence begins the decay from the T2 state, although the existing records do not indicate any clear experimental sign that would mark the T1 - T2 transition.

The experiment of Park \emph{et al.}  \cite{Barenghi-PL84-1981} used a wide square tube ($1 \times 1~\rm cm$) at relative high velocity; they found
a single superfluid state, more similar to the T1 state (as confirmed by Tough \emph{et al.} \cite{Martin-PRB27-1983}). This result would strongly suggest that in Schwarz and Rozen's experiments (rectangular section $1 \times 2.32~\rm cm$) the decay originated from the T1 state.

The problem is not only the lack of information about $L(0)$, $\gamma$ and the nature of the tangle which begins the decay, but also that
our model requires values for $c_L$ and $I_l$, which, like $\gamma$,  are provided only in calculations performed with a uniform normal fluid.  In conclusion, our model neglects the issue of the T1 or T2 state for lack of precise information (both from experiments
and from theory) and of consistent numbers to use.

In conclusion, we have reconsidered the experimental data of the decay of counterflow turbulence observed by Schwarz and Rozen \cite{Schwarz-Rozen} and by Skrbek \emph{et al.} \cite{Skrbek}, and compared them with our minimal two-fluid model of turbulence decay. The model correctly predicts a change in the nature of the vortex tangle at a time $t_{sat}$ which agrees fairly well even quantitatively, with the observed  onset of semiclassical $t^{-3/2}$ decay after an initial transient.
We have found that, contrary to common perception, the early observations of Schwarz and Rozen are in qualitative agreement with those of Skrbek \emph{et al}. Note that the $L$-dependence of $\beta$, ignored by Schwarz and Rozen\cite{Schwarz-Rozen} is essential in our model, because it is responsible for the self-induction of the vortex tangle in three dimensions, and hence for the more precise description of the decaying tangle.

Still, our model is perhaps to simple to accurately account for the long time (i.e., $t>t_{sat}$) asymptotic behavior $L\propto D t^{-3/2}$ that is displayed over at least an order of magnitude in various second sound experiments on decaying counterflow \cite{Smith-grid, Skrbek, QFS}. This power law behavior seems robust and allowed to determine the values of the effective kinematic viscosity $\nu_{\rm{eff}}$ \cite{TimKinVisc},  if the decay law is written in a classical-like form assuming that the Kolmogorov K41 form of the energy spectra over length scales up to $D$ (for further details we direct the reader to the review of Skrbek and Sreenivasan \cite{SkrSreeni}):
\begin{equation}
 L(t)=\frac {D(3C)^{3/2}}{2 \pi \kappa
\sqrt{\nu_{\rm{eff}}}}(t+t_{vo})^{-3/2}\cong\frac {D(3C)^{3/2}}{2
\pi \kappa \sqrt{\nu_{\rm{eff}}}}t^{-3/2}\,\,\,, \label{decay}
\end{equation}
where $C$ is the Kolmogorov constant and $t_{vo}$ stands for the virtual origin time that in most cases can be neglected. Values of  $\nu_{\rm{eff}}$ extracted this way from Prague experiments on decaying counterflow are in fair agreement with values obtained from the decaying grid turbulence in He II. It can be shown that our model would lead to the asymptotic behavior $L\propto D t^{-3/2}$ if $\beta \propto \sqrt{t}/D$ at late time, or, in terms of vortex line density $\beta \propto L^{-1/3} D^{-2/3}$ rather then $\beta \propto\ln{(L^{-1/2}/a_0)}$ as it was introduced by Schwarz. It is a challenge for the future to extend our model (which most likely will have to involve another length scale -- an outer scale of turbulence $D$ -- as in the classical case) in such a way that the experimentally observed robust asymptotic behavior will be described more accurately.

{\bf Acknowledgements:} M.S. thanks the Instituto Nazionale di Alta
Matematica  for supporting his research visit to Newcastle University.
C.F.B. and Y.A.S. are supported by the Leverhulme Trust. The research of LS is supported by the research plan MS 0021620834 of the Czech Republic and by GA\v{C}R 202/08/0276.


\end{document}